# The lens was fabricated by fluidic shaping

Chuanzhu Cheng[1], Fanru Kong[1], Yuqing Liu[1*]

*Center for Advanced Optoelectronic Functional Materials Research, and Key Laboratory for UV-Emitting Materials and Technology of Ministry of Education, National Demonstration Center for Experimental Physics Education, School of Physics, Northeast Normal University, 5268 Renmin Street, Changchun 130024, Jilin, China*

**Abstract:** As an important optical component, lens is widely used in scientific inquiry and production. At present, lens manufacturing mainly relies on grinding, polishing and other methods. However, these methods often require expensive equipment and complex processes. This paper presents a method of injecting liquid material into the frame structure and curing it quickly. At the same time, based on the principle of energy minimization, we give a set of theory that can accurately predict the lens face shape, and give the simulation results by software. In this paper, 3D printing technology was used to produce different shapes of borders, which were used to produce free-form surface and spherical lens samples. By characterizing their surface contours and optical properties, the practicability of the method was verified. This method has the advantages of low cost, fast forming, high surface smoothness, and can theoretically prepare any size aperture lens, which has great potential for development.
**Key words:** lens, fluidic shaping, 3D printing, Bessel function

## 1. Introduction

Lens is an important part of optical system, widely used in optical imaging, information transmission, illumination detection and other fields[1-7].Nowadays, the production of lenses mainly relies on single-point diamond cutting[8, 9], ultra-precision grinding[10], injection molding[11, 12] and other methods,however, these methods have high requirements for equipment and processes.In recent years, the rise of 3D printing technology has also made us see a new direction, using 3D printing technology, we can make almost any type of lens[13-15].3D printing mainly includes digital light processing (DLP), stereophotolithography (SLA), two-photon polymerization (TPP), etc. DLP technology is the use of digital micro-mirror device DMD to reflect the light from the light source to the photopolymer material to achieve the effect of layer curing, but this processing method has obvious "ladder effect," often requires complex post-processing processes[16,17].SLA technology uses a laser beam to illuminate the target range point by point, with the lifting platform in the vertical direction, to realize the processing of three-dimensional structures. Compared with DLP, SLA printing can obtain a higher quality surface[18]. Femtosecond laser is an advanced processing technology based on the principle of two-photon polymerization. It has the advantages of high precision and low thermal effect, and its processing accuracy is higher than that of traditional continuous laser[19]. Because of its high precision and single point processing, its processing efficiency is low and it is difficult to achieve mass production. Regardless of the processing method, the processing time is proportional to the lens volume, so it is not suitable for making large volume lenses.

Another method is to make use of the naturally smooth surface of polymer droplets to prepare high-quality surface lenses. Some researchers put PDMS droplets on the surface of the substrate, and through the upright and inverted substrate, under the action of surface tension, the polymer droplets solidified to obtain lenses with different curvatures [20]. However, this method has some limitations, as the volume of the droplet becomes larger, gravity begins to replace the surface tension, and the liquid on the upright substrate can no longer accumulate too high in the vertical direction, and the liquid on the inverted substrate may drip.In 2021, Valeri Frumkin and Moran Bercovici effectively eliminated the effect of gravity by submerging the lens liquid into another insoluble liquid, and using a frame structure to bind the lens liquid, and then curing the lens liquid to obtain a lens. This method is called fluidic shaping technology[21]. Fluidic shaping technology can produce large volume, high surface quality lenses in a short time, and only a frame structure is required as an experimental device, which greatly reduces the production cost. By changing the liquid volume, density difference, frame structure and other parameters of the lens, the lens surface shape can be adjusted within a certain range, so as to prepare a variety of lenses such as spherical, aspherical and free-form surfaces[22]. Based on the principle of free energy minimization, the expression of the lens surface under certain boundary conditions is given, and the simulation results are given by software.

On the basis of the previous researches, the systematic errors caused by the limitation of volume control accuracy of fluidic shaping technology are analyzed in detail, and the reason why this method is not suitable for the fabrication of small aperture lenses is explained. The spherical lens and free-form lens were prepared, and the surface profile and optical properties of the samples were tested, which proved the practicability of the method.

## 2.Solution and simulation of lens surface

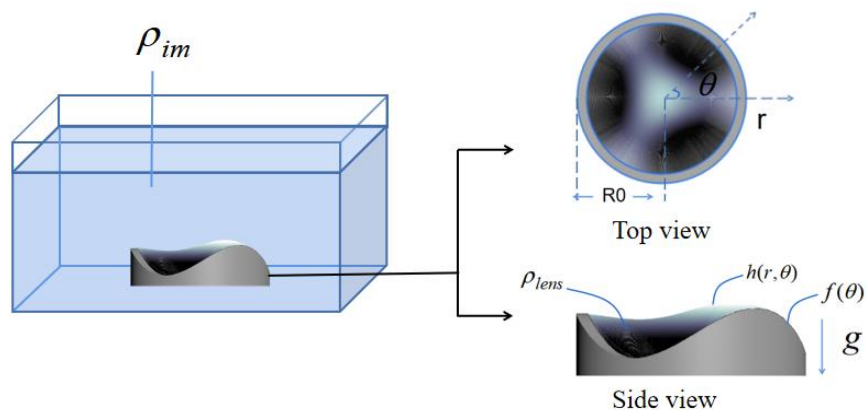

FIG. 1 Experimental device diagram

The experimental device is shown in Figure 1. The frame structure is placed in a sink filled with immersed liquid, and then a certain volume of lens liquid is injected into the frame structure, so that the lens liquid closely fits the inner wall of the frame structure. The two liquids are not mutually soluble, and a interface is formed at the interface, which is represented by h(r, θ). The density of the lens liquid is set to $\rho_{lens}$, the density of the immersed liquid is $\rho_{im}$, and the effective density is defined $\Delta\rho=\rho_{lens}-\rho_{im}$, when the lens liquid is in the immersed liquid and is affected by buoyancy. At the same time, the parameters are defined $Bo = \frac{|\Delta\rho|gR_0^2}{\gamma}$, where g is the acceleration of gravity, $\gamma$ is the surface tension coefficient, and the shape distribution of the boundary satisfies the function f(θ), So h(R₀,θ)=f(θ), at the boundary.

When the system reaches stability, under the constraint of a fixed volume, the surface shape of the lens liquid will be formed so that the free energy of the system will reach the lowest state. The expression of the system free energy function $E$ is as follows

$$E = \int_0^{2\pi} \int_0^{R_0} (\gamma\sqrt{1+(\frac{dh}{dr})^2+\frac{1}{r^2}(\frac{dh}{d\theta})^2} - \frac{1}{2}\Delta\rho gh^2 + \lambda h)r dr d\theta \qquad (2\text{-}1)$$

For the convenience of calculation, we define the function to be integrated in the free energy expression as the F function, that is

$$F(r,\theta) = (\gamma\sqrt{1+(\frac{dh}{dr})^2+\frac{1}{r^2}(\frac{dh}{d\theta})^2} - \frac{1}{2}\Delta\rho gh^2 + \lambda h)r \qquad (2\text{-}2)$$

When the system reaches stability, under the constraint of a fixed volume, the surface shape of the lens solution forms such that the free energy E of the system consists of two parts, the surface energy between the liquids and the effective gravitational potential energy, the last term represents the volume constraint, λ represents the Lagrange multiplier. When the whole system reaches a steady state, that is, when the energy of the whole system is the lowest, then the first derivative of the energy should be 0, that is

$$\delta E = 0 \qquad (2\text{-}3)$$

We get the Euler-Lagrange equation

$$\frac{\partial F}{\partial h} - \frac{d}{dr}\frac{\partial F}{\partial h_r} - \frac{d}{d\theta}\frac{\partial F}{\partial h_\theta} = 0 \qquad (2\text{-}4)$$

Where $h_r$ and $h_\theta$ represent the first derivative of h with respect to r and θ, respectively. Dimensionless parameters are defined as follows

$$R = \frac{r}{R_0}, H(R,\theta) = \frac{h(R)-h_0}{h_c}, P = (\frac{\lambda}{\Delta\rho gh_c} - \frac{h_0}{h_c}), \varepsilon = (\frac{h_c}{R_0})^2, x = R\sqrt{Bo}, h_0 = \frac{\int_0^{2\pi} f(\theta)}{2\pi}$$

$h_c$ stands for surface type variable.

Definition

$$\bar{H}(R,\theta) = H(R,\theta) - P \tag{2-5}$$

By separating the variables

$$\bar{H}(x,\theta) = \sum_{n=0}^{\infty} A_n J_n(x)\cos(n\theta) + \sum_{n=1}^{\infty} B_n J_n(x)\sin(n\theta) \tag{2-6}$$

Equation (2-6) is a general solution of a function related to H, where $J_n(x)$ represents an nth-order Bessel function with x as the independent variable. The corresponding particular solution can be obtained by applying special boundary conditions

$$f(\theta) = a_0 + \sum_{n=1}^{\infty} a_n \cos(n\theta) + \sum_{n=1}^{\infty} b_n \sin(n\theta) \tag{2-7}$$

The boundary conditions are used at the boundary $R=R_0$

$$h(R_0,\theta) = f(\theta) \tag{2-8}$$

The solution of $h$ under certain boundary conditions is obtained

$$h(r,\theta) = a_0 + P*(1 - \frac{J_0(\frac{r}{R_0}\sqrt{Bo})}{J_0(\sqrt{Bo})})$$

$$+ \sum_{n=1}^{\infty} (a_n \cos(n\theta) + b_n \sin(n\theta)) \frac{J_n(\frac{r}{R_0}\sqrt{Bo})}{J_n(\sqrt{Bo})} \tag{2-9}$$

The value of $P^*=P*h_c$ is given by the volume constraint

$$P^* = (a_0 - \frac{V}{\pi R_0^2}) \frac{J_0(\sqrt{Bo})}{J_2(\sqrt{Bo})} \tag{2-10}$$

According to the expression of face shape (2-9), the simulation results of part of face shape are given, and the influence of different factors on face shape is analyzed in detail, as shown in Figure 2.

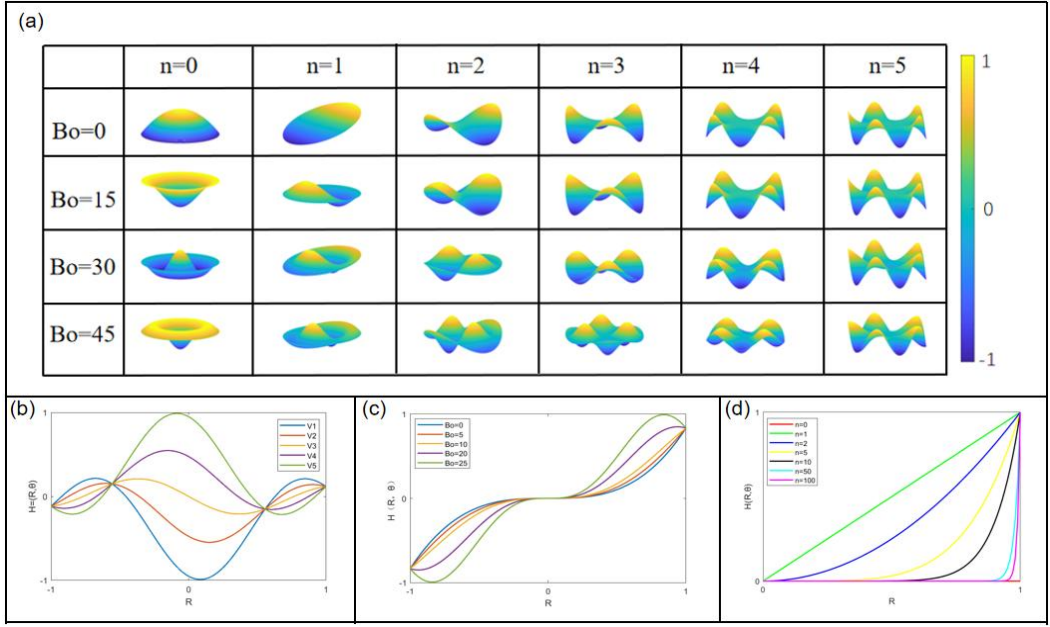

FIG. 2 Surface simulation results. (a) Simulation results under different experimental conditions; (b) The effect of liquid volume $V$ on the face type; (c) The effect of the $Bo$ face type; (d) The effect of the border structure on the face shape

Here, we discuss a special case, when $n=0$, the shape of the border is simplified from the original non-rotationally symmetric boundary to the hollow cylinder boundary, and the face type expression of the lens is also simplified to a certain extent, as follows

$$h(r,\theta) = a_0 + P^*\left(1 - \frac{J_0(\frac{r}{R_0}\sqrt{Bo})}{J_0(\sqrt{Bo})}\right) \qquad (2\text{-}11)$$

At this time, the parameter $Bo$, $P$ determines the surface shape of the lens, and the value of $P$, $Bo$ depends on the volume and density difference of the liquid. By changing the liquid volume and density difference of the lens, we can change the surface shape of the lens, and a series of rotationally symmetric lenses can be obtained. By changing the value of the volume $V$, we can obtain a series of rotationally symmetric Bessel lenses, as shown in Figure 3.

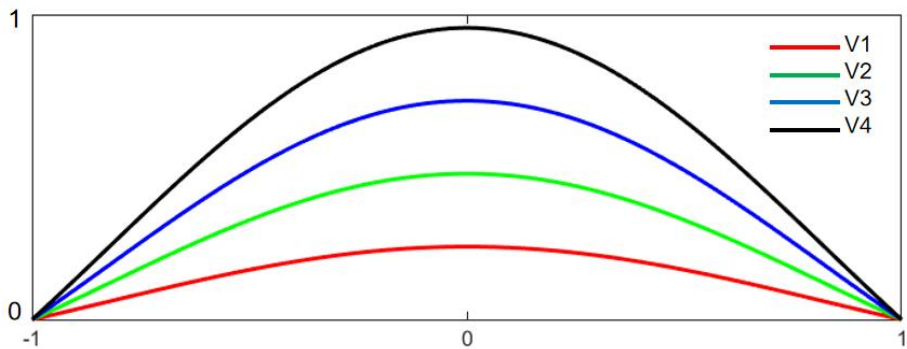

FIG. 3. When $Bo=5$, when $V$ takes a series of values within a certain range, lenses with different curvatures will be formed

Similarly, a series of lenses with different face shapes can also be obtained by changing the value of *Bo*, as shown in Figure 4.

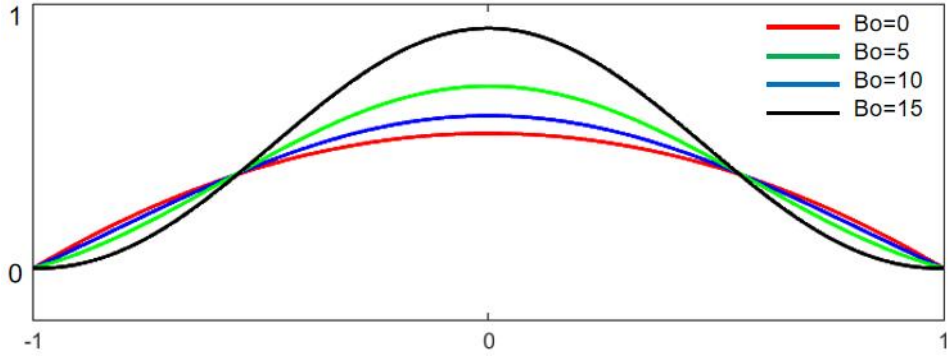

FIG. 4. Under the premise of constant volume, different face shapes will be formed when *Bo* is selected as 0,5,10,15 respectively

When *Bo*=0, that is, under the condition of neutral buoyancy, the surface shape appears as a spherical lens, and the surface shape function is further simplified.

$$h(r,\theta) = a_0 + 2(\frac{V}{\pi R_0^2} - a_0)(1 - R^2) \tag{2-12}$$

## 3. Fabrication and characterization of lenses

### 3.1 Spherical lens fabrication and error analysis

We built the laser direct writing system shown in Figure 5 and used it to process the border structure required by the experiment. The computer can control the deflection of the mirror and the switch of the optical gate by recognizing the information of the model file, and then control the processing position of the laser. The processing can be realized in space with the movement axis in the Z direction.

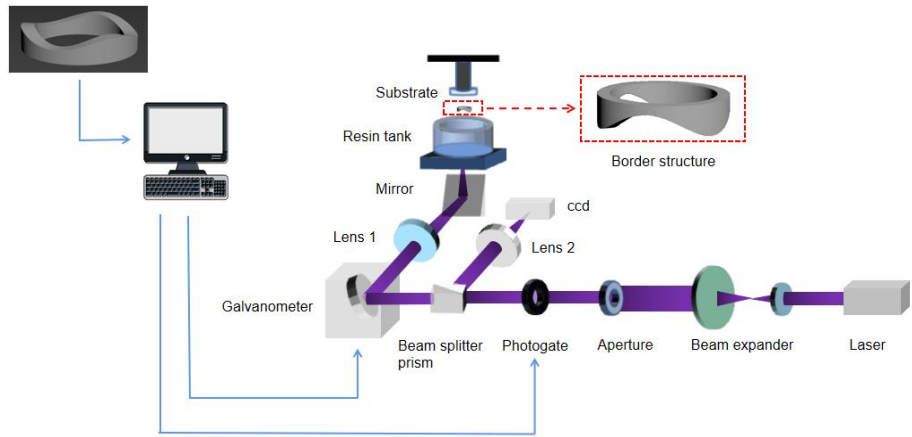

Figure 5 Schematic diagram of laser direct writing system

A spherical lens with a net diameter of 18.56 mm was designed and fabricated by this method, and the imaging effect was tested. a 5 mm aperture diaphragm was added to the optical path to reduce the influence of boundary spherical difference, as shown in FIG. 6 (a). According to the design parameters, we can see that the required border size is 9.28 mm in radius and 5 mm in height. In the theoretical part, we have given the expression of spherical lens. For a spherical lens with a certain aperture, the curvature of the surface becomes the only variable. The curvature formula at a certain point on the curve is as follows

$$k = \frac{|f''(t)|}{[1+f'(t)]^{3/2}} \tag{3-1}$$

By combining the formula (2-11), we get the relation between the radius of curvature of the spherical lens and the volume of the lens liquid

$$\rho = \frac{R_0^2}{4(a_0 - \frac{V}{\pi R_0^2})} \tag{3-2}$$

PDMS will shrink during the curing process. After consulting the merchant, it is learned that the density before curing is 0.98 g/cm³, and the density after curing is 1 g/cm³. The volume of liquid required can be deduced from the design parameters to 1.574 mL, considering the shrinkage effect of PDMS during curing, The liquid volume of the lens is 1.606 mL by means of inverse compensation. We added ethanol solution (density 0.789 g/cm³) to deionized water to regulate the density of the immersed solution. In this way, we can arbitrarily control the density of the mixture between 0.789 g/cm³ and 1.024 g/cm³ by adjusting the ratio. We established the frame structure model required by the experiment and processed it by laser 3D printing. Figure 6 (b) is the frame modeling, and Figure 6 (c) is the physical frame. In order to avoid lens fluid leakage, we use a transparent glass to seal the bottom end of the frame, add lens fluid to the frame, and add immersion liquid to soak the entire frame structure, in this process, we need to accurately control the liquid volume, so we use microfluidic injection pump (Suzhou Jinhao Microfluidic Technology Co., LTD., WH-SP-02) Inject 1.606 mL of liquid into the border. Then the whole device is put into the incubator, held at 60 °C for 4 h, until the material is completely cured and taken out, as shown in Figure 6 (d) for the spherical lens sample.

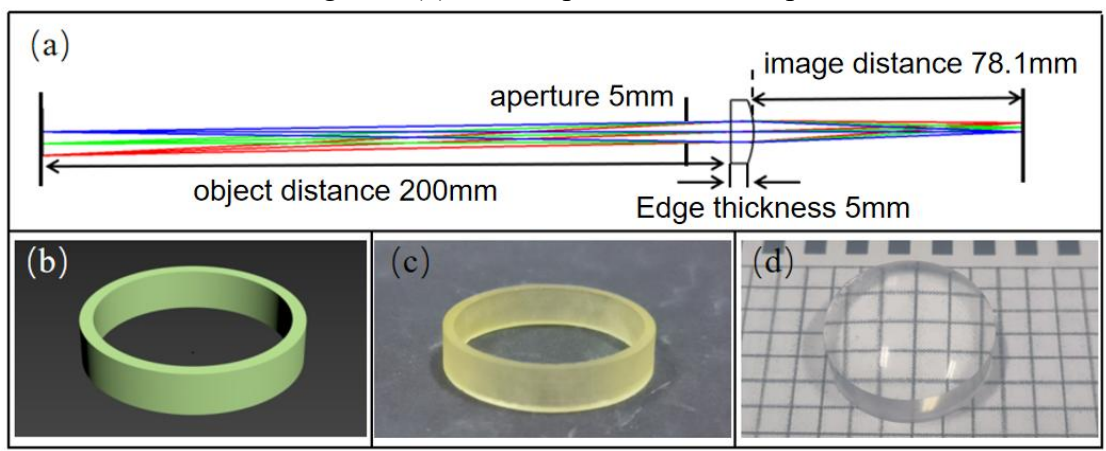

FIG. 6 (a) Design of 18.56 mm spherical lens; (b) Border model; (c) machined frame structures; (d) Spherical lens sample

The profile of the sample was characterized by confocal microscopy and the error analysis was compared with the design results. The test results showed that the maximum error was 6.1 μm, as shown in FIG.7 (a) (b), which was within the allowable error range. Atomic force microscopy was used to select a surface range of 5 μm×5 μm to test the surface roughness, and the test result was Rq=0.906 nm, as shown in Figure 7 (c).

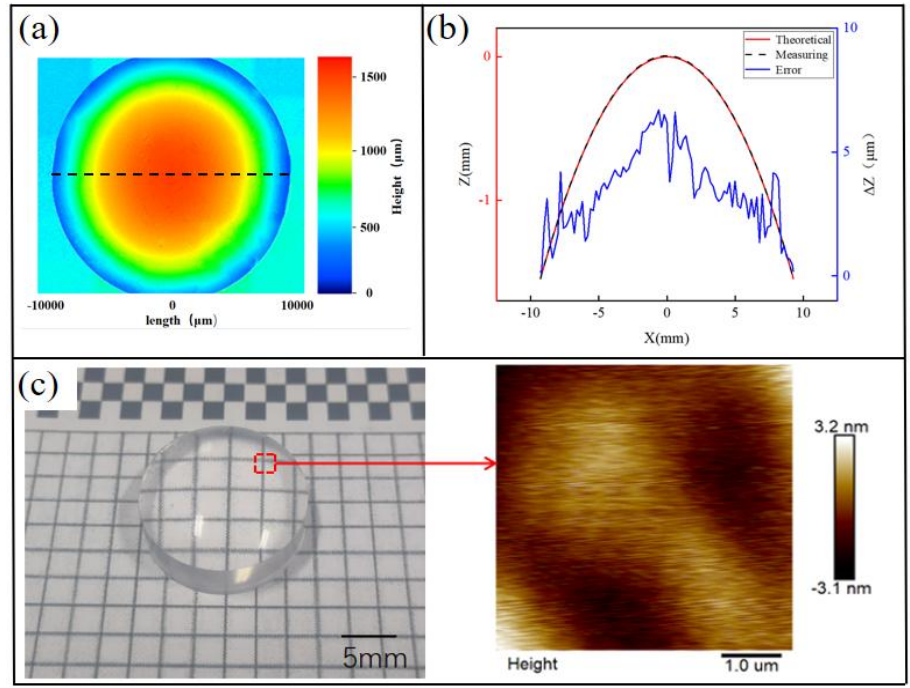

Figure 7. (a) confocal test results; (b) Profile error analysis; (c) Surface roughness testing

We built an optical path to test the imaging effect of the lens, as shown in Figure 8 (a). The target value of the MTF curve was calculated by identifying the pattern collected by the CCD. When the target value was 0.3, it corresponded to 64 1p/mm, which fitted well with the design result.

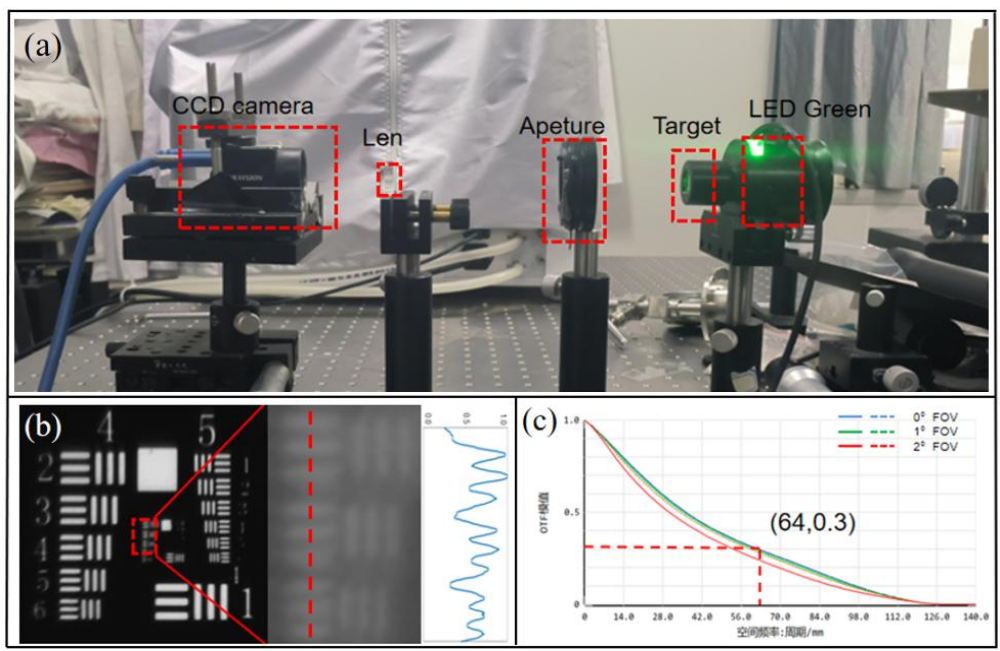

Figure 8 (a) Optical path diagram of imaging test; (b) Test results collected on CCD; (c) Spherical lens MTF design curve

## 3.2 fabrication and characterization of free-form lens

We have shown the fabrication and characterization of spherical lenses, and the method is not limited to the fabrication of simple spherical lenses. By modifying the boundary conditions of liquid contact, a free-form lens with complex surface profile can be obtained. A border with a diameter of 10mm and a full set of $f(\theta) = 0.4\sin(4\theta)$ distribution was constructed and machined. See Figure 9 (a) (b). The lens was prepared according to the process shown in Figure 9 (c), in which 0.2 mL of PDMS lens solution was injected and deionized water was used as the immersion solution. The final sample was shown in Figure 9 (d) (e).

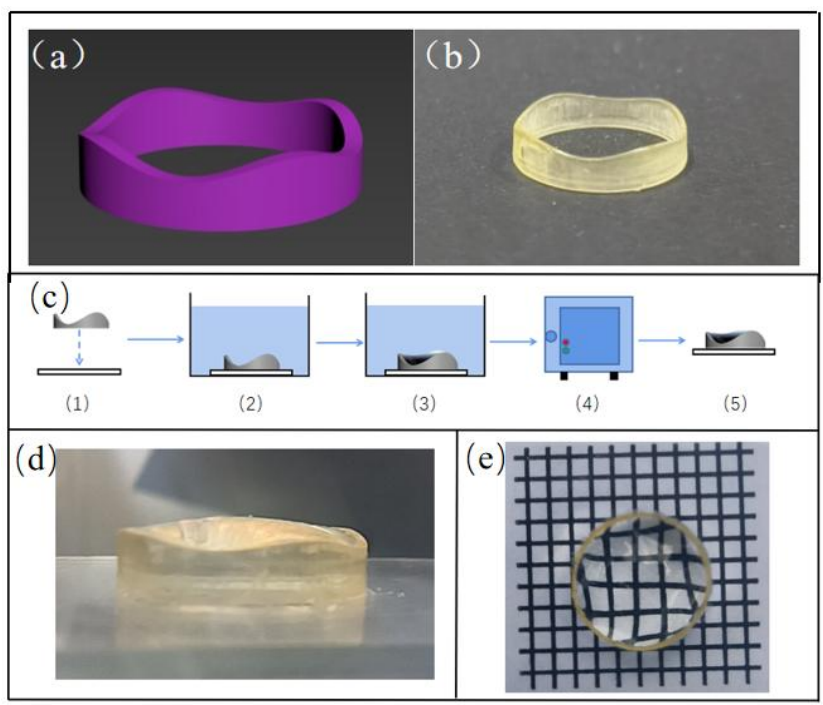

FIG. 9 (a) Frame structure model; (b) machined frame structures; (c) Experimental flow chart; (d) Side view of free-form surface sample; (e) Top view of free-form surface sample

Parametric conditions in the experiment can be used to give positive simulation results, as shown in FIG.10 (a). A confocal microscope was used to characterize the surface of the lens sample, as shown in Figure 10 (b). The three directional pairs of 0, π/4 and π/8 were selected to compare and analyze the contour errors between the experiment and the theory, and the results are shown in Figure 10 (c). The error of the contour lines in different directions is not exactly the same, but the maximum error is within 13.8 μm, which is within the allowable error range. Figure 10 (d) (e) (f) shows the comparative analysis results in three different angles and directions.

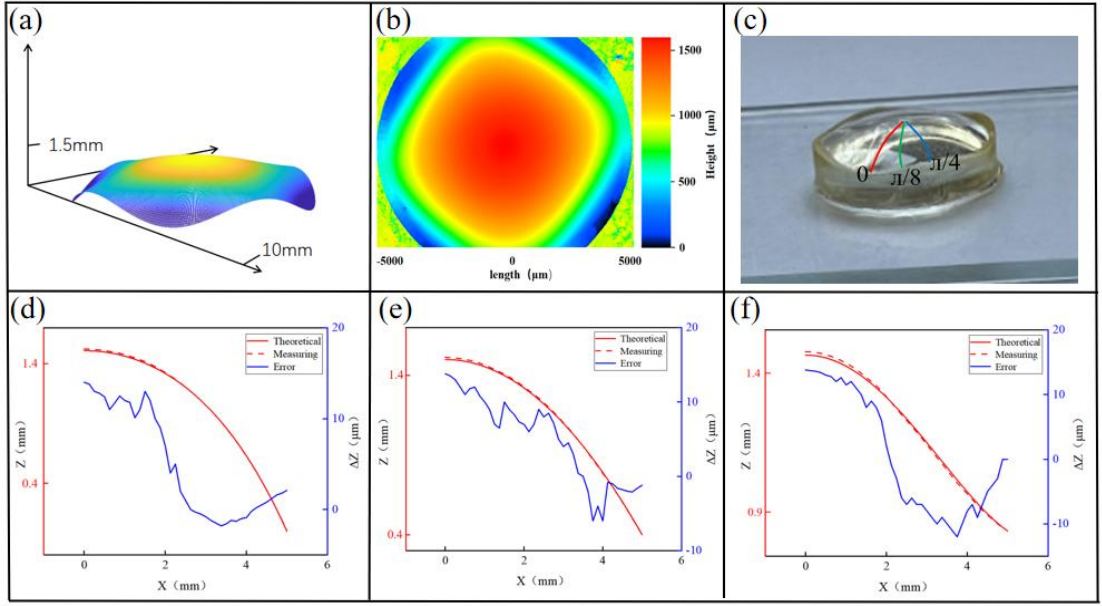

FIG. 10 (a) Surface simulation results; (b) confocal microscope rendering; (c) Contour diagrams in three directions: 0, π/8, π/4; (d) Comparison between simulation and experiment in 0° direction; (e) Comparison of simulation and experiment in π/8 direction; (f) Comparison between simulation and experiment in π/4 direction;

## 4.Conclusion

We introduce a method of making lenses using fluidic shaping technology, which can produce lenses with high quality surfaces in a short time. The shape of the surface is determined by the volume, density difference and the shape of the border, and because the contact surface between the liquids is naturally smooth, a very high surface quality can be obtained without complex post-treatment processes. The experimental process does not require any expensive experimental equipment, only two kinds of immiscible liquids and a frame structure, and there is no difference in time regardless of the size of the lens made, which greatly saves the economic cost and time cost. At the same time, from the perspective of free energy, a set of theoretical model is given which can simulate the lens surface under different experimental conditions, and it fits the experimental results well.

On the basis of the existing experiments, we can further increase the variables, so that the freedom of the surface can be further regulated. For example, by suspending the border, a lens with two faces can be produced, and the liquid type contacted by the upper and lower two faces can be adjusted, so as to achieve more complex regulation.